\documentstyle[12pt]{article}
\newcommand{\mibitem}[1]{\bibitem{#1}}

\newcommand{\be}{\begin{equation}}
\newcommand{\ee}{\end{equation}}
\newcommand{\ba}{\begin{eqnarray}}
\newcommand{\ea}{\end{eqnarray}}
\newcommand{\bastar}{\begin{eqnarray*}}
\newcommand{\eastar}{\end{eqnarray*}}
\newcommand{\half}{{1 \over 2}}

\newcommand{\tensor}{\otimes}

\newcommand{\M}{{\cal M}}
\newcommand{\N}{{\cal N}}

\newcommand{\X}{{\cal X}}

\newcommand{\G}{{\cal G}}

\newcommand{\cL}{{\cal L}}

\begin{document}
\begin{titlepage}
\begin{flushright}
UU-ITP 10/94 \\
hep-th/9406068
\end{flushright}

\vskip 0.7truecm

\begin{center}
{ \large \bf EQUIVARIANT MORSE THEORY \\ \vskip 0.2cm
AND QUANTUM INTEGRABILITY \\  }
\end{center}

\vskip 1.5cm

\begin{center}
{\bf Antti J. Niemi $^{*}$ } \\
\vskip 0.4cm
{\it Department of Theoretical Physics, Uppsala University
\\ P.O. Box 803, S-75108, Uppsala, Sweden \\}
\vskip 0.4cm
and \\
\vskip 0.4cm
{\bf Kaupo Palo $^{*}$ } \\
\vskip 0.4cm
{\it Department of Theoretical Physics, Uppsala University
\\ P.O. Box 803, S-75108, Uppsala, Sweden \\}
\vskip 0.0cm
and \\
\vskip 0.0cm
{\it Institute of Physics, Estonian Academy of Sciences \\
142 Riia St., 202 400 Tartu, Estonia }

\end{center}

\vskip 1.5cm

\rm
\noindent
We investigate an equivariant generalization of Morse theory for a
general
class of integrable models. In particular, we derive equivariant
versions of
the classical Poincar\'e-Hopf and Gauss-Bonnet-Chern theorems and
present the
corresponding path integral generalizations. Our approach is based on
equivariant cohomology and localization techniques, and is closely
related to
the formalism developed by Matthai and Quillen
in their approach to Gaussian shaped Thom forms.
\vfill

\begin{flushleft}
\rule{5.1 in}{.007 in}\\
$^{*}${\small E-mail: $~~$ {\small \bf
niemi@rhea.teorfys.uu.se} $~~~~$ {\bf
palo@rhea.teorfys.uu.se} \\ }
\end{flushleft}

\end{titlepage}

\vfill\eject

\baselineskip 0.65cm

\noindent
{\bf 1. Introduction}
\vskip 1.0cm

Equivariant cohomology is presently attracting much interest both in
Physics
and Mathematics.  This is largely due to its relevance to various
localization
formulas originally introduced by Duistermaat and Heckman, and its
connections
to cohomological topological field theories. More recently,
equivariant
cohomology has also found applications in the investigation of
$W$-gravity and
the formalism is also relevant in a geometric loop space approach to
Poincar\'e
supersymmetric theories.

The original localization formula by  Duistermaat and Heckman
\cite{mst}, \cite{dh} (for a review, see \cite{bgv}) concerns
exponential integrals over a $2n$ dimensional symplectic   manifold
$\M$ {\it i.e.} classical partition functions of the form
\be
Z ~=~ \frac{1}{(2\pi)^n} (i\phi)^n \int \frac{1}{n!} \omega^n
e^{i\phi H} ~=~
\frac{1}{(2\pi)^n} \int \exp \{ i\phi (H+\omega) \}
\label{ci}
\ee
where $\omega$ is the symplectic two-form. If the Hamiltonian $H$
determines a
global symplectic action of a  circle $S^1 \sim U(1)$ - or more
generally the
global action of a torus - on the manifold with isolated and
nondegenerate
fixed points $p$, the integral (\ref{ci}) localizes to these points,
\be
Z ~=~ \left( {2 \pi \over i \phi}
\right)^n \hskip -0.3cm \sum\limits_{dH=0  } \! \! \exp\{ i
\frac{\pi}{4} \eta_{H} \} { \sqrt{ \det || \omega_{\mu\nu} || } \over
\sqrt{
\det || \partial_{\mu\nu}
H || } } \exp\{ i \phi H \}
\label{dh}
\ee
Here $\eta_H$ is the $\eta$-invariant of the matrix
$\partial_{\mu\nu}H$ when
viewed as a linear operator on $TM_p$, {\it i.e.} if we denote the
dimensions
of the eigenspaces of the matrix $\partial_{\mu\nu}H$ at a
(non-degenerate)
critical point $p$ by $dim T_p^+$ and $dim T_p^-$,
\[
\eta_H ~=~ dim T_p^+ \ - \ dim T_p^-
\]

As explained by Berline and Vergne \cite{bv}, the integration formula
(\ref{dh})  can be
interpreted in the context of equivariant cohomology. In the case of
a circle
action, the pertinent equivariant cohomology is described by an
equivariant
exterior derivative
\[
d_H ~=~ d + i_H
\]
where $i_H$ is the (nilpotent) contraction operator along the
Hamiltonian vector field of
$H$. The operator $d_H$ squares to the Lie derivative of the circle
action,
\[
\cL_H ~=~ di_H + i_Hd
\]
which implies that it is nilpotent on the subcomplex $\Lambda_{H}$
of $U(1)$   invariant
exterior forms
\[
\cL_H \Lambda_{H} ~=~ 0
\]
Furthermore, since
\[
d_H (H + \omega) ~=~ 0
\]
we conclude that $H+\omega$ can be viewed as an equivariant extension
of the
symplectic two-form $\omega$. In particular, the integrand in
(\ref{ci}) is equivariantly closed and the integration formula
(\ref{dh}) can be seen as a consequence of an equivariant version of
Stokes theorem.

An infinite dimensional generalization of the Duistermaat-Heckman
formula was
presented by Atiyah and Witten \cite{aw}. They considered
loop space equivariant cohomology described by the loop space
differential
operator
\[
Q_{\dot z} ~=~ d + i_{\dot z}
\]
and they were interested in evaluating a supersymmetric path integral
that
describes the Atiyah-Singer index of a Dirac operator on a Riemannian
manifold.
The crucial idea in their work is that the fermionic bilinear in the
supersymmetric action can be interpreted as a loop space symplectic
two-form,
and integration over the fermions yields the loop space Liouville
measure.
Their approach was generalized by Bismut \cite{bis} to twisted
operators,   and to the
computation of the Lefschetz number of a Killing vector field acting
on the
manifold.

In \cite{osv}, \cite{jap} the action of four dimensional topological
Yang-Mills   theory was interpreted in terms of equivariant cohomology
and Weil algebra. In \cite{aj} Atiyah and Jeffrey gave an
interpretation of its partition function as a regularized Euler class
on a vector bundle associated to the bundle ${\cal A} \to {\cal
A}/{\cal G}$ of connections on a principal bundle. This interpretation
of cohomological topological field theories is an infinite dimensional
generalization \cite{aj}, \cite{mb} of the formalism developed earlier
by Matthai and   Quillen \cite{mq}, who explained how representatives
of the Thom class of   vector bundles can be constructed using
equivariant cohomology. They also applied this formalism to establish
a direct connection between the Poincar\'e-Hopf and Gauss-Bonnet-Chern
theorems in classical Morse theory using localization methods.

In a series of papers \cite{oma1}-\cite{oma8} the quantum mechanics of
circle actions of isometries on symplectic manifolds was considered
using a different method of loop space localization. The
derivation of the twisted Atiyah-Singer index theorem and its
generalizations was also considered in this formalism \cite{oma3},
\cite{oma4} and the ensuing equivariant cohomology and loop space
symplectic geometry was applied to formulate general
Poincar\'e-supersymmetric quantum field theories in a geometrical
framework \cite{oma3}, \cite{oma5}.

\vskip 0.3cm

In the present paper we shall be interested in applying the
equivariant
cohomology and localization techniques developed in
\cite{oma1}-\cite{oma8}  to   investigate
certain geometrical aspects of quantum integrability, for the general
class of
Hamiltonians that determine the global action of a circle on the
phase space.
In particular, we  explain how loop space equivariant cohomology can
be used to
construct novel quantum mechanical partition functions that are based
on this
family of Hamiltonians. We show how these partition functions can
be
evaluated exactly by localization methods. Our final results are
integrals of
equivariant characteristic classes over a moduli space which
describes the
classical configurations of the underlying dynamical system.

Our main results can be viewed as an equivariant version of
classical Morse
theory. In particular, we explain how loop space equivariant
generalizations of
the Poincar\'e-Hopf and Gauss-Bonnet-Chern theorems can be derived,
with the
equivariance determined by the Hamiltonian dynamics of $H$. Our
formalism is
closely  related to the Matthai-Quillen formalism, and in a sense it
can be
viewed as an "equivariantization" of their work.

In order to describe in more detail the results that we shall derive
here, we first recall some aspects of classical Morse theory (from the
present point of view see {\it e.g.} \cite{bgv}, \cite{mb},
\cite{bbrt}). For this we   consider critical points $p$ of a smooth
function $F$ - the Morse function -   on a
compact oriented $2n$ dimensional manifold $\cal M$,
\[
dF_{|_p} ~=~ 0
\]
If these critical points $p \in \cal M$ of $F$  are isolated and
non-degenerate, the Poincar\'e-Hopf theorem states that the Euler
characteristic ${\cal X}({\cal M})$ of $\cal M$ is related to these
critical points by
\be
{\cal X}({\cal M}) ~=~ \sum\limits_{i=0}^{2n} (-1)^i {\rm dim} \
H^i(\M;R) ~=~
\sum\limits_{dF=0} {\rm sign}( \det
|| { \partial^2 F (p) \over \partial x^\mu \partial x^\nu}  ||)
\label{ph}
\ee
with $x^\mu$ local coordinates around $p$. In particular, (\ref{ph})
means that the sum over the critical points of $F$ is a topological
invariant of the   manifold,
independently of the function $F$.

Here we shall be interested in generalizations of (\ref{ph}) to sums
that   are of the
form
\be
\sum\limits_{dF=0} {\rm sign}( \det || { \partial^2 F (p) \over
\partial   x^\mu
\partial x^\nu } ||) \cdot \exp\{ i \phi  F (p) \}
\label{eph}
\ee
where $\phi$ is a parameter.  We shall find that such generalizations
of (\ref{ph}) are
also related to an invariant of the manifold $\M$ which is an
equivariant version of the Euler class. Furthermore, we
shall explain how sums like (\ref{eph}) arise in the loop space, and we
also   present
the appropriate degenerate versions.

\vskip 0.3cm

In section 2. we shall introduce some background material on
equivariant
cohomology. In section 3. we present a derivation of the
Duistermaat-Heckman
integration formula (\ref{dh}) and generalize it to the degenerate
case.   This
derivation introduces techniques that we use in the subsequent
sections to
discuss more general localizations formulas.
In section 4. we introduce the supersymmetric complex $S^*\M$ and
explain how
the Matthai-Quillen formalism follows. We also generalize this
formalism by
equivariantizing it with respect to a Hamiltonian $H$ that generates
the action
of a circle on the phase space. In section 5. we use our formalism to
derive
equivariant generalizations of the Poincar\'e-Hopf and
Gauss-Bonnet-Chern
theorems in the finite dimensional context. In section 6. we return
to the
Duistermaat-Heckman integration formula, by
generalizing it to the loop space. The techniques we introduce in
this section
are then applied in the last two sections to localize loop space
integrals over
the supersymmetric complex $S^*\M$. First, in section 7. we
derive the Poincar\'e-Hopf and Gauss-Bonnet-Chern theorems for an
arbitrary,
non-gradient vector field.  In  section 8. we then extend these
theorems to
the equivariant context determined by a nontrivial Hamiltonian
dynamics.


\vskip 1.0cm
\noindent
{\bf 2. Equivariant cohomology on a symplectic manifold}

\vskip 0.7cm

In this section we introduce relevant concepts in
equivariant cohomology. We shall be interested in a $2n$ dimensional
compact
symplectic manifold  $\M$, with
local coordinates $x^\mu$ and Poisson bracket
\[
\{x^\mu , x^\nu \} ~=~ \omega^{\mu\nu}(x)
\]
Here $\omega^{\mu\nu}$ is the inverse matrix to the closed symplectic
two-form on $\M$,
\[
\omega ~=~ \half \omega_{\mu\nu} dx^\mu \wedge dx^\nu
\]
\[
d\omega ~=~ 0
\]
so that locally we can introduce a one-form $\vartheta$  called
the symplectic potential such that
\[
\omega ~=~ d\vartheta
\]

\vskip 0.3cm
We are interested in the equivariant cohomology $H^*_\G(\M)$
associated with the symplectic action of a Lie group
$\G$ on the manifold $\M$,
\[
\G \times \M ~ \to ~ \M
\]
If the action of $\G$ is free
{\it i.e.} the only element of $\G$ which acts trivially is the unit
element, the coset $\M/\G$ is well defined  and the $\G$-equivariant
cohomology of $\M$ coincides with the ordinary cohomology of the
coset,
\[
H^*_\G(\M) \ = \ H^*(\M/G)
\]
In the case of non-free $\G$-actions, more elaborated methods are
needed to compute the equivariant cohomology. Three different
approaches have been introduced to model $H^*_\G(\M)$, using
differential forms on $\M$ together with polynomial functions and
forms on the
Lie algebra $\bf g$ of $\G$. The two classical approaches are the
Cartan and
Weil models, and they are interpolated by the BRST model as described
{\it
e.g.} in \cite{oma8}, \cite{kalk}. In the following we
shall mainly need the Cartan model.

The $\G$-action on $\M$ is generated by vector fields $\X_\alpha$,
$\alpha =1,...,m$ that realize the commutation relations of the
Lie-algebra $\bf g$,
\[
[ \ \X_\alpha \ , \ \X_\beta \ ] ~=~ f^{\alpha\beta\gamma} \X_\gamma
\]
with $f^{\alpha\beta\gamma}$ the structure constants of $\bf g$. With
$\X$ a generic vector field on $\M$, we denote contraction along $\X$
by
$i_\X$. In particular, the basis of contractions corresponding to the
Lie algebra generators $\{ \X_\alpha \}$ is denoted by $i_{\X_\alpha}
\equiv i_\alpha$. The pertinent Lie-derivatives are
\[
\cL_{\alpha} ~=~ d i_\alpha + i_\alpha d
\]
where $d$ is the exterior derivative on the exterior algebra
$\Lambda(\M)$ of
the manifold $\M$. They generate the $\G$-action on $\Lambda(\M)$,
\[
[ \ \cL_\alpha  \ , \ \cL_\beta \ ] ~=~ f^{\alpha\beta\gamma}
\cL_\gamma
\]

We shall assume that the action of $\G$ is symplectic so that
\[
\cL_\alpha \omega ~\equiv ~ d i_\alpha \omega ~=~ 0 ~~~~~{\rm for ~
all ~}
\alpha
\]
Provided the one-forms $i_\alpha\omega$ are exact (for this the
triviality of $H^1(\M,R)$ is sufficient), we can then introduce the
corresponding momentum map
\[
H_\G: ~ \M \ \to {\bf g}^*
\]
where ${\bf g}^*$ is the dual Lie algebra. This yields a one-to-one
correspondence between the vector fields $\X_\alpha$ (and
corresponding
Lie-derivatives $\cL_\alpha$) and certain functions $H_\alpha$ on
$\M$, the
components of the momentum map
\[
H_\G ~=~ \phi^\alpha H_\alpha
\]
where $\{ \phi^\alpha \}$ is a (symmetric) basis of the dual Lie
algebra ${\bf
g}^*$, and
\[
i_\alpha \omega ~=~ - d H_\alpha
\]
or in local coordinates
\[
\X_\alpha ~=~ \omega^{\mu\nu} \partial_\mu H_\alpha \partial_\nu
\]

\vskip 0.3cm

In the sequel we shall only consider the simplest example of a group
action on the symplectic manifold $\M$, the action of a
circle $\G = U(1) = S^1$. It is determined by a momentum map $H$
corresponding
to a Hamiltonian vector field $\X_H$ as the only generator of the Lie
algebra
${\bf u}(1)$ of $U(1)$,
$$
\X_H ~=~ \omega^{\mu\nu}\partial_\mu H \partial_\nu
$$
We introduce the following equivariant exterior derivative
operator on $\M$,
\be
d_H ~=~ d + \phi i_H
\label{eed}
\ee
Here the factor $\phi$ is a real parameter that we interpret as a
generator of
the algebra of polynomials on ${\bf u}(1)$ {\it i.e.} as a generator
of the
symmetric algebra $S({\bf u}(1)^*)$ over the dual of the Lie algebra
of $U(1)$.
Thus the operator (\ref{eed}) acts on the complex $S({\bf u}(1)^*)
\tensor   \Lambda
(\M)$. Since
\[
d_H^2 ~=~ \phi ( d i_H + i_H d) ~=~ \phi \cL_H
\]
we conclude that on the $U(1)$ invariant subcomplex $( S({\bf
u}(1)^*) \tensor \Lambda (\M) )^{U(1)}$ the action of $d_H$ is
nilpotent and defines the $U(1)$-equivariant cohomology of $\M$ as
the
$d_H$-cohomology of $( S({\bf u}(1)^*) \tensor \Lambda (\M)
)^{U(1)}$.
Since the operations of evaluating $\phi$ and formation of cohomology
commute for abelian group actions the results coincide
independently of the interpretation of $\phi$, and for notational
simplicity we shall in the following
usually set $\phi = 1$. This model for the $U(1)$-equivariant
cohomology of
$\M$ is the abelian Cartan model.

\vskip 0.4cm
In the following we shall need a canonical realization of the various
operations on the algebra $S({\bf u}(1)^*) \tensor \Lambda (\M) $.
For
this we introduce canonical momentum variables $p_\mu$ which are
conjugate to
the coordinates $x^\mu$ of the original symplectic manifold, identify
the basis
of one-forms $dx^\mu$ with anticommuting $\eta^\mu$ and realize the
contraction
operator acting on $\eta^\mu$ canonically by $\bar\eta_\mu$, using
Poisson   brackets
\be
\{ p_\mu , x^\nu \} ~=~ \{ \bar\eta_\mu , \eta^\nu \} ~=~
\delta_\mu^\nu
\label{ecv}
\ee
In terms of these variables  the exterior derivative, contraction and
Lie derivative can be realized by the Poisson bracket actions of
\ba
d &=& p_\mu \eta^\mu
\nonumber \\
i_H &=& \X^\mu_H \bar\eta_\mu
\label{cld} \\
\cL_H &=& \X^\mu_H p_\mu + \eta^\mu \partial_\mu \X^\nu_H \bar\eta_\nu
\nonumber
\ea
\vskip 0.4cm
\noindent
Finally, since ($\phi = 1)$
\be
d_H (H + \omega ) ~=~  dH + i_H \omega ~=~ 0
\label{eqvh}
\ee
we conclude that $H + \omega$ is an element of
${H^*}_{U(1)}(\M)$, that is it  determines an equivariant cohomology
class. This is an equivalence class consisting of elements in
$\Lambda(\M)$ which are linear combinations of zero- and two-forms
and
can be represented as
\[
H + \omega + d_H \psi
\]
where $\psi$ is in $\Lambda_H(\M)$ {\it i.e.} it satisfies
\[
\cL_H \psi ~=~ 0
\label{fof}
\]


\vskip 1.0cm
\noindent
{\bf 3. Duistermaat-Heckman Integration Formula}
\vskip 0.6cm

The integration formula by Duistermaat and Heckman concerns the exact
evaluation of the classical partition function
\be
Z ~=~ \int \omega^n e^{i\phi H}
\label{cpf}
\ee
where $H$ is a hamiltonian function that determines the global
symplectic
action of $S^1 \sim U(1)$ on the phase space $\M$, and in physical
applications
$\phi$ is identified as the inverse temperature. The integration
measure is the
phase space Liouville measure which is a canonically invariant
measure on $\M$.

If the critical points of $H$ are isolated and nondegenerate, the
integration
formula by Duistermaat and Heckman states that (\ref{cpf}) localizes
to the
critical points of $H$,
\be
Z~=~ \sum\limits_{dH=0}  {
\sqrt{det||\omega_{\mu\nu} ||}\over \sqrt{ det|| \partial_{\mu\nu}  H
|| } } \ \exp\{ i\phi H\}
\label{cdh}
\ee
where for simplicity we have included the phase factor that appears
in (\ref{dh}) to
the definition of the determinant of $H$, and we have also absorbed
the
additional factors that appear in (\ref{dh}) to the normalization of
$Z$.

In this section we shall present a derivation of the integration
formula (\ref{cdh})
and explain how it generalizes to the case where the critical points
of $H$ are
degenerate. In the subsequent sections we then use the techniques
that we
introduce here to derive new integration formulas for integrals that
are more
general than (\ref{cpf}).

Using the fact that integration picks up the $2n$-form, modulo an
overall
normalization we can write (\ref{cpf}) as
\be
Z ~=~  \int \exp\{i \phi(H+\omega) \}
\label{cpf2}
\ee
or in local coordinates
\[
Z ~=~ \int dx d\eta  \exp \{ i \phi ( H + \half \eta^{\mu}
\omega_{\mu\nu}
\eta^\nu \}
\]
In order to prove the integration formula (\ref{cdh}), we introduce the
following generalization of (\ref{cpf2})
\be
Z_\lambda ~=~ \int  \exp\{ i\phi (H + \omega) + \lambda d_H
\psi \}
\label{cpf3}
\ee
Here $\psi$ is a one-form and $\lambda$ is a parameter. We shall
first
argue that if $\psi$ is in the $H$-invariant subspace,
\be
\cL_H \psi ~=~ 0
\label{ldc}
\ee
the integral (\ref{cpf3}) does not depend on $\lambda$. Since the
integrals (\ref{cpf2}) and (\ref{cpf3})
coincide for $\lambda \to 0$, this implies that these integrals
coincide for
all values of $\lambda$. By evaluating (\ref{cpf3}) in the $\lambda
\to   \infty$ limit we
then obtain the integration formula (\ref{cdh}).

Notice in particular, that the $\lambda$-independence of (\ref{cpf3})
means    that
(\ref{cpf2}) {\it only} depends on the equivalence class that
$H + \omega$ determines in the equivariant cohomology
${H^*}_{U(1)}(\M)$.

In order to establish the $\lambda$-independence of (\ref{cpf3}), we
consider   an
infinitesimal variation $\lambda \to \lambda + \delta
\lambda$ and show that
\[
Z_\lambda ~=~ Z_{\lambda + \delta \lambda}
\]
For this we introduce the following infinitesimal change of
variables in (\ref{cpf3}):
\ba
x^\mu ~\to~  x^\mu ~+~ \delta x^\mu & = & x^\mu ~+~
\delta \psi \cdot d_H x^\mu  =  x^\mu ~+~ \delta \psi \eta^\mu
\nonumber \\
\eta^\mu ~\to~  \eta^\mu ~+~ \delta \eta^\mu & = & \eta^\mu ~+~
\delta \psi
\cdot d_H \eta^\mu  =  \eta^\mu ~-~ \delta \psi \X^\mu_H
\label{cov}
\ea
where
\[
\delta \psi ~=~ \delta \lambda \cdot \psi
\]
As a consequence of (\ref{eqvh}) and (\ref{ldc}) the exponential in
(\ref{cpf3}) is invariant under the change of variables (\ref{cov}).
However, the Jacobian is nontrivial:
$$
dx d \eta ~\to~ (1+d_{H}\delta \psi )d x d \eta ~\sim~
\exp\{ d_{H}(\delta \psi)\}dx  d\eta ~=~ \exp\{  \delta \lambda d_H
\psi
\} dx d \eta
$$
Hence
\[
Z_\lambda ~=~ \int dx d\eta \exp\{ i \phi (H + \omega) + \lambda d_H
\psi + \delta \lambda d_H\psi \} ~=~ Z_{\lambda + \delta \lambda}
\]
and the classical partition function (\ref{cpf2}) depends
{\it only} on the equivalence class that $H + \omega$ determines in
the
equivariant cohomology ${H^*}_{U(1)}(\M)$.

The $\lambda$-independence of (\ref{cpf3}) implies (\ref{cdh}): For
this we   first
observe that since the group $U(1)$ is compact we may construct a
metric tensor
$g_{\mu\nu}$ on $\M$ for which the canonical flow of $H$ is an
isometry,
\be
\cL_H g ~=~ 0
\label{ldm}
\ee
or in components,
\[
\X^\rho_H\partial_\rho g_{\mu\nu} + g_{\mu\rho} \partial_\nu
\X^\rho_H +
g_{\nu\rho} \partial_\mu \X^\rho_H ~=~ 0
\]
Such a metric is obtained by selecting an arbitrary
Riemannian metric  $g_{\mu\nu}$ on $\M$, and averaging it over the
group
$U(1)$. A converse  is also true: Since $\M$ is compact the isometry
group of
$g_{\mu\nu}$ must be compact.

We select
\be
\psi ~=~ i_H g ~=~ g_{\mu\nu} \X^\mu_H \eta^\nu
\label{psi}
\ee
As a consequence of (\ref{ldm}),
\[
\cL_H \psi ~=~ 0
\]
and
\[
d_H \psi ~=~ K + \Omega
\]
and the integral
\be
Z ~=~ \int \exp\{ i \phi (H + \omega) -
\frac{\lambda}{2}(K+\Omega) \}
\label{zko}
\ee
is independent of $\lambda$. Here we have defined
\[
K ~=~ g_{\mu\nu} \X^\mu_H \X^\nu_H
\]
and
\be
\Omega_{\mu\nu} ~=~ - \Omega_{\nu\mu} ~=~ \frac{1}{2}  [ \partial_\mu
(g_{\nu\rho} \X^\rho_H) -  \partial_\nu (g_{\mu\rho} \X^\rho_H) ]
{}~=~
\frac{1}{2}   [ \nabla_\mu (g_{\nu\rho} \X^\rho ) -  \nabla_\nu
(g_{\mu\rho}
\X^\rho) ]
\label{rmm}
\ee
which is called the Riemannian momentum map \cite{bgv}. Here
$\nabla_\mu$ is   the
covariant derivative
$$
\nabla_\mu (g_{\nu\rho}\X^\rho) ~=~ \partial_\mu (g_{\nu\rho}\X^\rho)
-
\Gamma^\sigma_{\mu\nu}
(g_{\sigma\rho}\X^\rho)
$$
and $\Gamma^\rho_{\mu\nu}$ is the Levi-Civita connection
$$
\Gamma^\rho_{\mu\nu} ~=~ \frac{1}{2} g^{\rho\sigma} ( \partial_\mu
g_{\nu\sigma} + \partial_\nu
g_{\mu\sigma} - \partial_\sigma g_{\mu\nu} )
$$
and the last relation in (\ref{rmm}) follows from antisymmetry.

We note that since $\Omega$ determines a closed two-form on $\M$ it
can be
viewed as a (degenerate) symplectic two-form. Furthermore, we find
that
$(H,\omega)$ and $(K,\Omega)$ defines a bi-hamiltonian pair in the
sense that
their classical trajectories coincide,
\[
\Omega_{\mu\nu} \dot{x^\nu} ~=~ \partial_\mu K ~=~ \Omega_{\mu\nu}
\omega^{\nu\rho} \partial_\rho H
\]
which is consistent with the classical integrability of $H$.

Explicitly, (\ref{zko}) is
\be
Z ~=~ \int dx d\eta \exp\{ i \phi ( H + \omega) - \frac{\lambda}{2}
g_{\mu\nu}\X^\mu_H\X^\nu_H
-  \frac{\lambda}{2}  \Omega_{\mu\nu} \eta^\mu \eta^\nu \}
\label{ezko}
\ee
and the integration formula (\ref{cdh}) follows
immediately when we recall that
\[
\delta (\alpha  x) \ = \ \frac{1}{| \alpha |} \delta(x) \ = \
\lim\limits_{\lambda \to \infty} \sqrt{ \frac{\lambda}{2\pi}}
e^{-\frac{\lambda}{2} (\alpha x)^2 }
\]
which localizes (\ref{ezko}) onto (\ref{cdh}),
$$
Z ~=~ \int dx d\eta \frac{ {\rm Pf} ( \Omega_{\mu\nu} ) }{\sqrt{\det
||g_{\mu\nu}||} }~\delta (\X_H ) ~ e^{i\phi (H + \omega)}
$$
\[
=~ \sum\limits_{dH=0}   { \sqrt{\det ||
\omega_{\mu\nu} ||} \over \sqrt{\det || \partial_{\mu\nu} H || }
}~\exp\{ i
\phi H\}
\]
Here we have used
\[
\partial_\mu \X^\nu_H ~=~ \omega^{\nu\sigma}\partial_\mu
\partial_\sigma H
\]
on the critical points $dH=0$ and included the phase factor that
appears in (\ref{dh}) to the definition of the determinant of
$\partial_{\mu\nu}H$.

\vskip 0.3cm

We now generalize (\ref{cdh}) for Hamiltonians $H$ with degenerate
critical   points. We
denote the critical submanifold of $H$ in $\M$ by $\M_{0}$ and by
$\N_{\bot}$
its normal bundle in $\M$. In a neighborhood near $\M_0$ we write the
local
coordinates $x^\mu$ as
\[
x^\mu ~=~ {\hat x}^\mu  + \delta x^\mu
\]
where $\hat x^\mu$ are local coordinates in $\M_0$,
\be
\X^\mu_H (\hat x ) ~=~ 0
\label{ccs1}
\ee
and $\delta x^\mu $ are local coordinates in $\N_{\bot}$.
Similarly, we introduce
$$
\eta^\mu ~=~ {\hat \eta}^\mu + \delta \eta^\mu
$$
where ${\hat \eta}^\mu$ are one-forms in $T^*\M_0$ and $\delta
\eta^\mu$ are
one-forms in $T^*\N_{\bot}$. In particular, the ${\hat\eta}^\mu$
satisfy
\be
\Omega_{\mu\nu}(\hat x) {\hat \eta}^\nu ~=~ 0
\label{ccs2}
\ee
for all $\mu$.

For large $\lambda$ the integral (\ref{ezko}) localizes exponentially
to the   vicinity of
$\M_0$, and consequently we can extend integration over all values of
$\delta
x^\mu$.  We introduce the following change of variables,
\ba
x^\mu & = & {\hat x}^\mu + \delta x^\mu ~ \longrightarrow ~ {\hat
x}^\mu +
\frac{1}{\sqrt{\lambda}} \delta x^\mu
\nonumber \\
\eta^\mu & = & {\hat \eta}^\mu + \delta \eta^\mu ~ \longrightarrow ~
{\hat
\eta}^\mu + \frac{1}{\sqrt{\lambda}} \delta \eta^\mu
\label{cov1}
\ea
Since the Jacobians for the $\delta x^\mu$ and for the $\delta
\eta^\mu$
cancel, the corresponding Jacobian is trivial.

Consider the last term in (\ref{ezko}),
\be
\frac{\lambda}{2} \eta^\mu \Omega_{\mu\nu} \eta^\nu
\label{kim1}
\ee
If we perform the change of variables (\ref{cov1}) and expand
(\ref{kim1}) in powers of $\sqrt{\lambda}$ we find that the term which
is proportional to   $\lambda$
\[
{\hat \eta}^\mu \Omega_{\mu\nu} {\hat \eta}^\nu
\]
vanishes by (\ref{ccs2}). Similarly, the terms which are proportional
to $\sqrt{\lambda}$
$$
{\hat \eta}^\mu \delta x^\rho \partial_\rho \Omega_{\mu\nu} \eta^\nu
+ \delta
\eta^\mu \Omega_{\mu\nu}
{\hat \eta}^\nu + {\hat \eta}^\mu \Omega_{\mu\nu} \delta \eta^\nu
$$
vanish as a consequence of (\ref{ccs1}), (\ref{ccs2}) and the Lie
derivative condition (\ref{ldm}) for the metric tensor.

We shall now proceed to investigate the ${\cal O}(1)$
contribution in this expansion,
\[
\frac{1}{2} {\hat \eta}^\mu {\hat \eta}^\nu \delta x^\rho \delta
x^\sigma
\partial_\rho
\partial_\sigma \Omega_{\mu\nu} + \delta \eta^\mu \delta x^\rho
\partial_\rho
\Omega_{\mu\nu}
{\hat \eta}^\nu + {\hat \eta}^\mu \delta x^\rho  \partial_\rho
\Omega_{\mu\nu}
\delta \eta^\nu
+ \delta \eta^\mu  \Omega_{\mu\nu} \delta \eta^\nu
\]
A lengthy but straightforward calculation  reveals, that these terms
can be
combined into
\be
\frac{1}{2} \delta \eta^\mu \Omega_{\mu\nu} \delta \eta^\nu +
\frac{1}{2} \delta x^\mu \Omega_\mu^\kappa R_{\kappa \nu \rho
\sigma}{\hat
\eta}^\rho {\hat \eta}^\sigma
\delta x^\nu
\label{kim2}
\ee
where
\[
{R^\rho}_{\sigma\mu\nu} ~=~ \partial_\mu \Gamma^\rho_{\nu\sigma} -
\partial_\nu \Gamma^\rho_{\mu\sigma} + \Gamma^\rho_{\mu\kappa}
\Gamma^\kappa_{\nu\sigma} - \Gamma^\rho_{\nu\kappa}
\Gamma^\kappa_{\mu\sigma}
\]
is the Riemann tensor. Hence in the $\lambda \to \infty$ limit
(\ref{kim1}) reduces to (\ref{kim2}).

Similarly, we expand the second term in (\ref{ezko}), {\it i.e.}
$$
\frac{\lambda}{2} g_{\mu\nu} \X^\mu_H \X^\nu_H
$$
in powers of $\sqrt{\lambda}$, after we first perform the change of
variables (\ref{cov1}). We find, that in the $\lambda \to \infty$ limit
the only term that survives is
\be
\frac{\lambda}{2} g_{\mu\nu} \X^\mu_H \X^\nu_H ~\longrightarrow~
\frac{1}{2} \delta x^\mu \Omega_\mu^\rho \Omega_{\rho\nu} \delta
x^\nu
\label{kim3}
\ee
{}From (\ref{kim1}) and (\ref{kim2}) we conclude that in the $\lambda \to
\infty$ limit the integrals over $\delta x^\mu$  and $\delta \eta^\mu$
in (\ref{ezko}) become Gaussian, and evaluating these integrals we get
\[
Z ~=~ \int\limits_{\M_0} dx d\eta \ exp \{ i \phi (H + \omega) \}
\cdot {  {\rm Pf } ( \Omega_{\mu\nu} )
\over \sqrt{ \det || \Omega_{\mu\sigma} ( \Omega^\sigma_\nu +
{R^\sigma}_{\nu
\rho \kappa } \eta^\rho \eta^\kappa ) || } }
\]

\be
=~ \int\limits_{\M_0} d x d\eta \ { \exp \{ i \phi (H + \omega) \}
\over {\rm Pf} ( \Omega_{\mu\nu} + R_{\mu\nu\rho\sigma} \eta^\rho
\eta^\sigma ) }
\label{dcdh}
\ee
where the $\det$ and $\rm Pf$ are evaluated over the normal bundle
$\N_\perp$.
This result generalizes (\ref{cdh}) to  the degenerate case. Indeed, if
we take the limit where $\M_0$ becomes a set of isolated and
nondegenerate critical
points and carefully account for the sign of the Pfaffian in
(\ref{dcdh}), we   find that
in this limit (\ref{dcdh}) reproduces (\ref{cdh}).

If we set $H=0$ in (\ref{dcdh}) we recognize in the numerator the
Chern
class of the symplectic two-form $\omega$ and in the denominator the
Euler
class of the curvature two-form $R_{\mu\nu}$. With nonzero $H$, we
can then
identify \cite{bgv}
\[
{\rm Ch} (H + \omega) ~=~ \exp \{ i \phi (H + \omega) \}
\]
as the {\it equivariant Chern class} of $\M_0$ and
\[
{\rm E}  ( \Omega_{\mu\nu} + R_{\mu\nu\rho\sigma} \eta^\rho
\eta^\sigma ) ~=~
{\rm Pf}( \Omega_{\mu\nu} + R_{\mu\nu\rho\sigma}
\eta^\rho \eta^\sigma)
\]
as the {\it equivariant Euler class} of $\M_0$, evaluated over the
normal
bundle $\N_\perp$. Equivariant characteristic classes \cite{bgv} are
generalizations
of ordinary characteristic classes to the equivariant context, and
provide
representatives of equivariant cohomology. In particular, we
note that
equivariant characteristic classes are independent of the connection
which implies that (\ref{dcdh}) is independent of the metric tensor
consistently with our general arguments.


\vskip 1.0cm
\noindent
{\bf 4. $\bf S^*M$ and generalization of the Matthai-Quillen
formalism}
\vskip 0.7cm
We shall now proceed to explain how the previous construction can be
extended
to derive more general integration formulas. For this, instead of
integrals
over the cotangent bundle $\M\tensor T^*\M$ with local coordinates
$x^\mu$ and
$\eta^\mu$ we shall in the following consider integrals that are
defined over
all four variables $x^\mu$, $\eta^\mu$, $p_\mu$, $\bar \eta_\mu$ that
we have
introduced in (\ref{ecv}).

Originally, we introduced $p_\mu$ as a canonical realization of the
local basis
for the tangent bundle $T\M$ and $\bar \eta_\mu$ as a canonical
realization of
the local basis of the contraction dual bundle of $T^*\M$. In the
following we
shall instead interpret these variables as follows: We view $x^\mu$
and $\bar
\eta_\mu$ as local coordinates on a supermanifold that we denote by
$S^*\M$.
In analogy with (\ref{cld}), we interpret $\eta^\mu$ as (part of) a
local   basis for the
cotangent bundle of $S^*\M$ with the identification $\eta^\mu \sim
dx^\mu$.
However, instead of viewing $p_\mu$ as a local basis for the tangent
bundle
$TS^*\M$ we shall in the following interpret $p_\mu$ as (part of) a
local basis
for the cotangent bundle of $S^*\M$ with the identification $p_\mu
\sim d \bar
\eta_\mu$.  The pertinent exterior derivative is
\be
d ~=~ \eta^\mu {\partial \over \partial x^\mu} + p_\mu {\partial
\over \partial
\bar \eta_\mu}
\label{esed1}
\ee
and it is a nilpotent operator on the exterior algebra
$\Lambda(S^*\M)$.

We also introduce a basis $i_\mu$, $\pi^\mu$ of contractions on the
exterior
algebra $\Lambda(S^*\M)$, dual to the basis $\eta^\mu$, $p_\mu$ of
one-forms,
\[
i_\mu \eta^\nu ~=~ \pi^\nu p_\mu ~=~ \delta^\nu_\mu
\]

In order to construct the Matthai-Quillen formalism we consider a
conjugation
of $d$ by a functional $\Phi$,
\be
d ~ {\buildrel {\Phi} \over {\longrightarrow} } ~
e^{-\Phi} d e^{\Phi} ~=~ d + [d , \Phi] + \frac{1}{2\!} [ [d , \Phi ]
, \Phi ]
+ ...
\label{esed2}
\ee
Since this conjugation is an invertible transformation, the
cohomologies of (\ref{esed1})
and (\ref{esed2}) coincide. Using the Levi-Civita connection
$\Gamma^\rho_{\mu\nu}$ we
define
\be
\Phi ~=~ - \Gamma^\rho_{\mu\nu} \pi^\mu \eta^\nu \bar \eta_\rho
\label{cfphi}
\ee
which yields for the conjugated exterior derivative
\be
d ~=~ \eta^\mu {\partial \over \partial x^\mu} + (p_\mu +
\Gamma^\rho_{\mu\nu}
\eta^\nu
\bar \eta_\rho) {\partial \over \partial \bar \eta_\mu} + (
\Gamma^\rho_{\mu\nu} p_\rho c^\nu -
\frac{1}{2}{R^\rho}_{\mu\sigma\nu} \eta^\nu \eta^\sigma \bar
\eta_\rho) \pi^\mu
\label{esed3}
\ee
In particular, we have
\ba
d x^\mu & = &  \eta^\mu \nonumber \\
d\eta^\mu & = & 0 \nonumber \\
dp_\mu & = & \Gamma^\rho_{\mu\nu} p_\rho \eta^\nu -
\frac{1}{2}{R^\rho}_{\mu\sigma\nu}
\eta^\nu \eta^\sigma \bar \eta_\rho \nonumber \\
d\bar \eta_\mu & =& p_\mu + \Gamma^\rho_{\mu\nu} \eta^\nu \bar
\eta_\rho \nonumber
\ea
where we identify the transformation laws of the standard ($N=1$)
deRham
supersymmetric quantum mechanics, with $p_\mu$ the auxiliary field
(see {\it e.g.} \cite{bbrt}). As explained in \cite{mb}, this means
that (\ref{esed3}) can be related to the   Matthai-Quillen formalism
\cite{mq}.
In particular, the corresponding quantum   mechanical path integral
determines an infinite dimensional version of the   Matthai-Quillen
formalism \cite{mb} and reduces to the original, finite dimensional
Matthai-Quillen
formalism by a regularization procedure. As explained in \cite{mb} we
can   use such a
path integral and localization methods to derive the
Gauss-Bonnet-Chern and
Poincar\'e-Hopf theorems of classical Morse theory. We shall not
reproduce
these derivations here, but refer to \cite{mb} for details.
Instead we shall now proceed to generalize the previous construction
of the Matthai-Quillen formalism to include
the action of a nontrivial vector field on
$S^*M$, which corresponds to the global circle action on the
original symplectic manifold $\M$ generated by our Hamiltonian
$H$.

We first recall the canonical realization (\ref{cld}) of the Lie
derivative along the Hamiltonian vector field $\X_H$ on the exterior
algebra $\Lambda(\M)$,
\[
\cL_H ~=~ \X^\mu_H p_\mu + \eta^\mu \partial_\mu \X^\nu_H \bar
\eta_\nu
\]
Notice that since it acts by the Poisson brackets (\ref{ecv}),
it is in fact defined on the exterior algebra $\Lambda(S^*\M)$.
In particular, on our canonical variables the Poisson bracket
action
of ${\cal L}_H$ is
\ba
\cL_H x^\mu & = & \X^\mu_H \nonumber \\
\cL_H \eta^\mu & = & \eta^\nu \partial_\nu \X^\mu_H
\nonumber \\
\cL_H \bar\eta_\mu & = & - \partial_\mu \X^\nu_H \ \bar\eta_\nu
\nonumber \\
\cL_H p_\mu & =& - \partial_\mu \X^\nu_H \ p_\nu - \eta^\nu
\partial_{\mu\nu}\X^\rho_H \
\bar\eta_\rho
\label{ldtl}
\ea
In order to generalize the Matthai-Quillen formalism, as a first step
we realize this action by a Lie derivative that acts as a
differential operator on the exterior algebra $\Lambda(S^*\M)$. For
this we define the following vector field on $TS^*\M$
\[
\X ~=~ \X^\nu_H \frac{\partial}{\partial x^\nu} - \bar\eta_\nu
\partial_\mu \X_H^\mu \frac{\partial}{\partial \bar \eta_\nu }
\]
and introduce the corresponding nilpotent contraction operator
on  $\Lambda(S^*\M)$,
\[
i_\X ~=~ \X^\mu_H i_\mu - \bar\eta_\nu \partial_\mu \X^\mu_H \pi^\nu
\]
and the corresponding equivariant exterior derivative
\be
Q_\X ~=~ d ~+~ i_\X ~=~ \eta^\mu {\partial \over \partial x^\mu} +
p_\mu
{\partial \over \partial \bar \eta_\mu} ~+~\X^\mu_H i_\mu -
\bar\eta_\nu
\partial_\mu \X^\mu_H \pi^\nu
\label{qchi}
\ee
We find that the ensuing Lie-derivative
\[
\L_\X ~=~ d i_\X \ + \ i_\X d
\]
\be
=~ \X^\mu_H \frac{\partial}{\partial x^\mu} +  \eta^\nu
\partial_\nu \X^\mu_H i_\mu
- \partial_\mu \X^\nu_H \bar\eta_\nu \frac{\partial}{\partial
\bar\eta_\mu}  - (\partial_\mu \X^\nu_H p_\nu  + \eta^\nu
\partial_{\mu\nu}
\X^\rho_H
\bar\eta_\rho ) \pi^\mu
\label{esld1}
\ee
then reproduces the transformation laws (\ref{ldtl}) on the variables
($x^\mu,   \
\bar\eta_\mu , \ \eta^\mu , \ p_\mu$).

Next, we introduce the conjugation (\ref{esed2}) for $Q_\X$ using the
functional $\Phi$ defined in (\ref{cfphi}). This yields for the
equivariant exterior derivative (\ref{qchi})
\[
Q_\X ~\to~ e^{\Phi} Q_\X e^{\Phi} ~=~ \eta^\mu \frac{\partial}{\partial
x^\mu} + (p_\mu + \Gamma^\rho_{\mu\nu}
\eta^\nu \bar\eta_\rho) \frac{\partial}{\partial \bar\eta_\mu} +
\X^\mu_H
i_\mu +
\]
\[
+ ( \Gamma^\rho_{\mu\nu}p_\rho \eta^\nu -
\frac{1}{2} {R^\rho}_{\mu\sigma\nu}\eta^\nu \eta^\sigma \bar\eta_\rho
-
\X^\nu_H \Gamma^\rho_{\mu\nu} \bar\eta_\rho -
\partial_\mu
\X^\nu_H \bar\eta_\nu ) \pi^\mu
\]
and for the Lie-derivative (\ref{esld1}) we find
\[
\L_\X ~=~   \X^\mu_H  \frac{\partial}{\partial x^\mu}
+ \eta^\nu
\partial_\nu \X^\mu_H i_\mu
- \partial_\mu \X^\nu_H \bar\eta_\nu
\frac{\partial}{\partial \bar\eta_\mu}  - p_\nu \partial_\mu
\X^\nu_H \pi^\mu
\]
\be
- \eta^\nu \bar\eta_\rho ( \X^\sigma_H
\partial_\sigma\Gamma^\rho_{\mu\nu} +
\partial_\nu\X^\sigma_H
\Gamma^\rho_{\mu\sigma}
+ \partial_\mu \X^\sigma_H \Gamma^\rho_{\sigma\nu} -
\Gamma^\sigma_{\mu\nu}
\partial_\sigma \X^\rho_H +
\partial_{\mu\nu} \X^\rho_H ) \pi^\mu
\label{esld2}
\ee
Here we recognize in the last term
\[
\X^\sigma_H \partial_\sigma\Gamma^\rho_{\mu\nu} +
\partial_\nu\X^\sigma_H
\Gamma^\rho_{\mu\sigma}
+ \partial_\mu \X^\sigma_H \Gamma^\rho_{\sigma\nu} -
\Gamma^\sigma_{\mu\nu}
\partial_\sigma \X^\rho_H +
\partial_{\mu\nu} \X^\rho_H
\]
the Lie-derivative of the connection one-form $\Gamma_\mu$ in the
{\it
original} manifold $\M$ along the Hamiltonian vector field $\X_H$.
Since $\X_H$ generates the global action of a circle on $\M$ which
leaves the
metric tensor $g_{\mu\nu}$ invariant, we conclude that on $\M$ we can
set
\be
\cL_H \Gamma_\mu ~=~ 0
\label{ldcon}
\ee
Consequently we find that (\ref{esld2}) simplifies into
\be
\L_\X ~=~   \X^\mu_H  \frac{\partial}{\partial x^\mu}
+ \eta^\nu
\partial_\nu \X^\mu_H i_\mu
- \partial_\mu \X^\nu_H \bar\eta_\nu
\frac{\partial}{\partial \bar\eta_\mu}  - p_\nu \partial_\mu
\X^\nu_H \pi^\mu
\label{esld3}
\ee
and instead of (\ref{ldtl}) we have the following transformation laws
for the local variables on $S^*\M$,
\ba
\L_\X x^\mu & = & \X^\mu_H \nonumber \\
\L_\X \bar\eta_\mu & = & - \partial_\mu \X^\nu_H \bar\eta_\nu
\nonumber \\
\L_\X \eta^\mu & = & \eta^\nu \partial_\nu \X^\mu_H
\nonumber \\
\L_\X p_\mu & = & -  p_\nu \partial_\mu \X^\nu_H
\label{eldtl}
\ea
In particular, comparing (\ref{ldtl}) and (\ref{eldtl}) we observe that
in (\ref{ldtl}) the first   three
have the appropriate covariant forms for the transformation of a
coordinate
$(x^\mu)$, one-form $(\eta^\mu)$ and its dual $(\bar \eta_\mu)$ on
$\M$
respectively under a coordinate transformation generated by the
vector field
$\X_H$, and the inhomogeneity in the last term reflects the fact that
under a
general coordinate transformation a
derivative transforms inhomogeneously.  However, in (\ref{eldtl}) {\it
all   four}
tansformations are generally covariant. In particular, the last one
has the
appropriate homogeneous form of the generally covariant
transformation law for
a covariant derivative on $\M$.

Obviously, the formalism that we have developed here can be viewed as
a generalization - or equivariantization - of the Matthai-Quillen
formalism \cite{mq}, \cite{mb}, and we now proceed to apply it to
derive new integration formulas.


\vskip 1.0cm
\noindent
{\bf 5. Equivariant Morse theory}
\vskip 0.7cm

We shall first apply our generalization of the Matthai-Quillen
formalism to
derive a new finite dimensional integration formula for integrals on
$\Lambda(S^*\M)$ that are of the form
\be
Z ~=~ \int dx dp d\eta d\bar\eta \exp \{ i\phi (H+\omega) + Q_\X \psi
\}
\label{esz}
\ee
Here the integration measure is the Liouville measure in the extended
phase
space $(x^\mu, p_\mu, \eta^\mu , \bar\eta_\mu)$ with Poisson brackets
(\ref{ecv}).
This is an invariant integration measure on $\Lambda(S^*\M)$. The
Hamiltonian
$H$ and the symplectic two-form $\omega$ are as before, {\it i.e.}
defined in
the original phase space $\M$ so that $H$ depends only on $x^\mu$
while
$\omega$ is a function of $x^\mu$ and a bilinear in $\eta^\mu$. The
equivariant
exterior derivative $Q_\X$ is
defined in (\ref{qchi})  and $\psi$ is an element in the subspace of
$\Lambda(S^*\M)$
that satisfies
\be
\L_\X \psi ~=~ 0
\label{lesldc}
\ee
where $\L_\X$ is the Lie derivative defined in (\ref{esld3}). In
particular,   since (\ref{esld3})
assumes (\ref{ldcon}) {\it i.e.} that the Lie derivative of the
Levi-Civita   connection
one-form $\Gamma_\mu$ on $\M$ vanishes, we again take the Hamiltonian
$H$ to be
a canonical generator of a global circle action on the original phase
space
$\M$.

Since
$$
Q_\X (H + \omega) ~=~ 0
$$
we conclude using our earlier arguments that if $\psi$ is in the
subspace (\ref{esldc}) the integral  (\ref{esz}) is invariant under
such local variations of $\psi$ that are in this subspace. Hence the
integral (\ref{esz})  only depends  on the equivariant cohomology
classes of $Q_\X$. However, since $H$ and $\omega$ do not   depend on
the variables $p_\mu$ and $\bar \eta_\mu$ we can not naively set
$\psi \to 0$
since in this limit the integral (\ref{esz}) is not properly defined.
This integral is
properly defined {\it only} if the $Q_\X \psi$-term depends
appropriately also on the variables $p_\mu$ and $\bar\eta_\mu$. Thus
we expect
that (\ref{esz}) does not coincide
with the Duistermaat-Heckman integral (\ref{cdh}); the equivariant
cohomology   on
$\Lambda(\M)$ does  not describe the equivariant cohomology on
$\Lambda(S^*\M)$.

In order to evaluate (\ref{esz}), we need to construct an appropriate
functional $\psi$. For this, we observe that since the basic
variables transform in the homogeneous and generally covariant manner
(\ref{eldtl}) under
the Lie derivative,  {\it any generally covariant quantity
which is built from $p_\mu,$, $\eta^\mu$, $\bar\eta_{\mu}$ and
invariant tensors on $\M$ such as the metric tensor $g_{\mu\nu}$,
the Hamiltonian vector field $\X^\mu_H$, the symplectic two-form
$\omega_{\mu\nu}$, the covariant derivative $\nabla_\mu$ {\rm etc.}
automatically satisfies the Lie-derivative condition (\ref{esldc})
on } $S^*\M$.
Notice that this is in a marked contrast with the Duistermaat-Heckman
case, where we have no general rule for the construction of
functionals $\psi$ that satisfy the condition (\ref{ldc}) on $\M$.

We shall first assume that the critical points of $H$ are isolated
and nondegenerate. We select
\be
\psi ~=~ \frac{1}{2} \X^\mu_H \bar\eta_\mu
\label{esldc}
\ee
As a generally covariant quantity, this automatically satisfies the
condition (\ref{esldc}). Explicitly,
$$
Q_\X \psi ~=~ p_\mu \X^\mu_H + \eta^\mu \nabla_\mu \X^\nu_H
\bar\eta_\nu
$$
where $\nabla_\mu$ denotes the covariant derivative with respect to
the
$H$-invariant Levi-Civita connection $\Gamma_\mu$ on $\M$.
Substituting in the
integral (\ref{esz}) we get
\[
Z ~=~ \int dx dp d\eta d\bar\eta \exp \{ i \phi (H + \omega) + p_\mu
\X^\mu_H + \eta^\mu \nabla_\mu \X^\nu_H \bar\eta_\nu \}
\]
We evaluate the $p_\mu$ integrals and the integrals over the
anticommuting
variables $\eta^\mu$ and $\bar\eta_\mu$. The result is
\be
Z ~=~ \int dx e^{i\phi H} \delta(\X^\mu_H)
\det || \partial_\mu \X^\nu_H || ~=~ \sum\limits_{dH = 0 } e^{-i \phi
H} {\rm sign}( \det
|| \partial_{\mu\nu} H  || )
\label{eph2}
\ee
where on the {\it r.h.s.} we recognize an equivariant version of the
quantity
that appears in the Poincar\'e-Hopf theorem.

Next we consider
\[
\psi ~=~ g^{\mu\nu} p_\mu \bar\eta_\nu
\]
As a generally covariant quantity, this again satisfies the condition
(\ref{esldc}) and we
have
\[
Q_\X \psi ~=~ g^{\mu\nu} p_\mu p_\nu - \frac{1}{2}
{R^\rho}_{\mu\sigma\nu}\eta^\nu \eta^\sigma \bar\eta_\rho \
g^{\mu\kappa}\bar\eta_\kappa -
\nabla_\nu \X^\mu_H \ g^{\nu\rho} \bar\eta_\mu \bar\eta_\rho
\]
When we substitute this in (\ref{esz}), evaluate the Gaussian integral
over   $p_\mu$ and
the integral over the anticommuting $\bar\eta_\mu$
we get
\be
Z ~=~ \int\limits_\M d x d\eta \ \exp\{ i \phi (H + \omega) \} \cdot
{\rm Pf}[
\nabla_\nu \X^\mu_H + \frac{1}{2} {R^\mu}_{\nu\rho\sigma} \eta^\rho
\eta^\sigma
]
\label{egb2}
\ee
We identify this as an equivariant version of
the quantity that appears in the Gauss-Bonnet-Chern theorem.

Combining (\ref{eph2}) and (\ref{egb2}) we obtain the following
equivariant   generalization of the
familiar relation between the Poincar\'e-Hopf and Gauss-Bonnet-Chern
theorems,
\be
\sum\limits_{dH = 0 } \!  e^{-i \phi H} {\rm sign}( \det
|| \partial_{\mu\nu} H  || ) = \int\limits_{\M} \! \ d x d\eta e^{  i
\phi (H +
\omega) }  {\rm Pf}[ \nabla_\nu \X^\mu_H + \frac{1}{2}
{R^\mu}_{\nu\rho\sigma}
\eta^\rho \eta^\sigma ]
\label{ephgb}
\ee
In particular, in the $\phi \to 0$ limit (\ref{ephgb}) reduces to the
standard relation
\[
\sum\limits_{dH = 0 } {\rm sign}( \det
|| \partial_{\mu\nu} H  || ) ~=~ \int\limits_{\M} \! \ d x d\eta \
{\rm Pf}[
\frac{1}{2} {R^\mu}_{\nu\rho\sigma} \eta^\rho \eta^\sigma ]
\]
between the Poincar\'e-Hopf and Gauss-Bonnet-Chern theorems.
As we have explained in the previous section, in this limit we
also reproduce the finite
dimensional Matthai-Quillen
formalism. (Notice that in taking the $\phi\to 0$ limit in the {\it
r.h.s.} of
(\ref{ephgb}) we have used the fact, that the integral picks up the
top-form of   the
Pfaffian.)

In order to generalize to the case where the critical point set
$\M_0$ of the
Hamiltonian $H$ is degenerate we first observe that both quantities
in the {\it
r.h.s.} of (\ref{ephgb}) are equivariantly closed on $\Lambda(\M)$,
\[
d_H \exp \{ i \phi(H +\omega) \} ~=~ d_H {\rm Pf}[
\nabla_\nu \X^\mu_H + \frac{1}{2} {R^\mu}_{\nu\rho\sigma} \eta^\rho
\eta^\sigma
] ~=~ 0
\]
Indeed, as we have explained in section 3. these quantities are the
equivariant
generalizations of the Chern class and Euler class on $\M$,
respectively. Using our standard arguments  we then conclude that the
following generalization of the integral in (\ref{ephgb}) over $\M$,
\be
Z ~=~ \int d x d\eta \  \exp\{ i \phi (H + \omega) + d_H \psi \}
\cdot {\rm Pf} [ \nabla_\nu \X^\mu_H +
\frac{1}{2} {R^\mu}_{\nu\rho\sigma} \eta^\rho \eta^\sigma ]
\label{ephgb2}
\ee
is independent of the functional $\psi$  and coincides with
(\ref{ephgb}),   provided
$\psi$ satisfies the Lie-derivative condition
\[
{\cal L}_H \psi ~=~ 0
\]
on $\Lambda(\M)$. If we
select ( \ref{psi}) and repeat the computation that yielded
the   degenerate
version (\ref{dcdh}) of the Duistermaat-Heckman integration formula, we
find   that the
contribution from the $d_H\psi$ -term in (\ref{ephgb2}) cancels the
Pfaffian in   (\ref{ephgb2})
except for the contribution that comes from the evaluation of the
determinant
over the normal bundle $\N_{\bot}$ of the critical submanifold $\M_0$
of $H$.
Consequently (\ref{ephgb2}) reduces to the following
integral over the critical submanifold $\M_0$ of the Hamiltonian $H$,
\[
Z ~=~ \int\limits_{\M_0} d x  d \eta \exp \{ i \phi (H + \omega) \}
\cdot {\rm Pf} [  \nabla_\nu \X^\mu_H + \frac{1}{2}
{R^\mu}_{\nu\rho\sigma}
\eta^\rho \eta^\sigma ]
\]
which can be viewed as an equivariant version of the
Gauss-Bonnet-Chern formula in the degenerate case.


\vskip 1.0cm
\noindent
{\bf 6. Duistermaat-Heckman formula in the loop space}
\vskip 0.7cm

We shall now proceed to generalize the previous results to loop
space {\it i.e.} path integrals. We again start by first considering
the loop
space generalization of the Duistermaat-Heckman integration formula,
and in the
subsequent sections we continue to more general loop space
integration
formulas.

In the Duistermaat-Heckman case we are interested in evaluating the
standard path integral
\[
Z ~=~
\int [dx] \sqrt{ \det || \omega_{\mu\nu} || } \exp\{ i
\int\limits_{0}^{T} \vartheta_{\mu} {\dot x}^{\mu} - H \}
\]
\be
=~ \int [dx] [d\eta]
\exp\{ i \int\limits_{0}^{T} \vartheta_{\mu} {\dot x}^{\mu} - H +
\frac{1}{2} \eta^\mu \omega_{\mu\nu} \eta^\nu \}
\label{dhls}
\ee
Here we have represented the path integral Liouville measure
using anticommuting variables $\eta^\mu$. We again assume that the
Hamiltonian
$H$ generates the action of a circle
$S^1$ on the classical phase space $\M$.

In order to evaluate (\ref{dhls}),  we interpret \cite{oma1} it as an
integral in the loop space $L\M$ over the phase space $\M$. This loop
space is   parametrized by the
time evolution $x^\mu \to x^\mu (t)$ with $x^\mu (0) = x^\mu (T)$. As
before,
we identify $\eta^\mu (t)$ as a basis of loop space one-forms,
$\eta^\mu \sim
dx^\mu $ and define exterior derivative on $L\M$ by lifting the
exterior
derivative from $\M$,
\be
d ~=~ \int\limits_0^T \! dt \ \eta^{\mu}(t) { \delta \over \delta
x^\mu (t)}
\label{lsedd}
\ee
In the following all time integrals will be implicit, and {\it e.g.}
instead of
(\ref{lsedd}) we write simply
\[
d ~=~ \eta^\mu \partial_\mu
\]
Similarly various other quantities on $\M$ can be lifted to $L\M$.
For example, by defining the loop space symplectic two-form
\[
\Omega_{\mu\nu}(t,t') ~=~ \omega_{\mu\nu} \delta(t-t')
\]
we have a symplectic structure on $L\M$.

We interpret the bosonic part of the action in (\ref{dhls}) as a
Hamiltonian   functional
on the loop space. The corresponding loop space hamiltonian vector
field  has
components
\be
{\cal X}_{S}^{\mu} ~=~ \dot x^\mu - \omega^{\mu\nu}
\partial_{\nu} H ~=~ \dot x^\mu - \X^\mu_H
\label{lsvf}
\ee
In particular, its critical points coincide with the classical
trajectories.

We introduce a basis $i_{\mu}$ of loop space contractions
dual to the basis of one-forms $\eta^\mu(t)$,
\[
i_{\mu}(t)\eta^{\nu}(t') ~=~ \delta_{\mu}^{\nu}(t-t')
\]
and define the loop space equivariant exterior derivative along
the loop space vector field (\ref{lsvf}),
\be
d_{S} ~=~ d+ i_{S} ~=~ \eta^\mu \partial_{\mu} + {\cal X}_{S}^{\mu}
i_{\mu}
\label{lsed}
\ee
The corresponding Lie-derivative is
\be
{\cal L}_{S} ~=~ d i_{S} + i_{S} d ~=~ \X^\mu_S \partial_\mu +
\eta^\mu \partial_\mu \X^\nu_S i_\nu
\label{lsld}
\ee
We introduce the invariant subcomplex of loop space exterior
algebra where the Lie derivative (\ref{lsld}) vanishes. In this
subspace (\ref{lsed})   is
then nilpotent and determines loop space equivariant
cohomology. In particular, the action in (\ref{dhls}) is equivariantly
closed,
\[
d_{S} (\vartheta_{\mu} {\dot x}^{\mu} -  H + \frac{1}{2} \eta^\mu
\omega_{\mu\nu} \eta^\nu ) ~=~ 0
\]
and determines an element of the corresponding equivariant cohomology
class.

Again, we conclude \cite{oma1} that if we add to the exponential in
(\ref{dhls}) a $d_{S}$ exact
term,
$$
S ~\to~ \int \vartheta_{\mu} {\dot x}^{\mu} - H + \frac{1}{2}
\eta^\mu \omega_{\mu\nu} \eta^\nu + d_{S} \psi
\]
where $\psi$ is in the nilpotent subspace
\be
{\cal L}_{S} \psi ~=~ 0
\label{ldls}
\ee
the corresponding path integral
\be
Z_{\Psi} ~=~ \int [dx] [d\eta] \ \exp \{ i \int
\vartheta_{\mu} {\dot x}^{\mu} - H + \frac{1}{2}
\eta^\mu \omega_{\mu\nu} \eta^\nu + d_{S}
\psi \}
\label{dhls2}
\ee
is invariant under local variations of $\psi$ and coincides with
(\ref{dhls}). By selecting  $\psi$ properly the path integral can then
be evaluated by the localization method.

In order to enumerate the different possibilities, we consider the
following
example,
\be
Z ~=~ \int [d \cos ( \theta )] [d \phi] \exp \{ i \int\limits_{0}^T
j \cos ( \theta ) \dot\varphi - j \cos (\theta) \}
\label{su2i}
\ee
This path integral is defined on the sphere $S^2$ and yields the
character of
$SU(2)$ in the spin-$j+\frac{1}{2}$ representation \cite{afs}.

The loop space Hamiltonian vector field (\ref{lsvf}) of the action in
(\ref{su2i}) has   two
components,
\ba
\X_\theta &=&  \dot\theta \nonumber \\
\X_\varphi &=& \dot\varphi - 1
\label{su2e}
\ea
For $T\not= 2\pi n $ the only $T$-periodic critical trajectories of
(\ref{su2e}) coincide with the critical points of the Hamiltonian $H =
j \cos   (\theta)$,
\[
\theta ~=~ 0, ~ \pi
\]
Consequently for $T\not= 2\pi n$ the critical point set of the action
in (\ref{su2i}) is
isolated and nondegenerate. On the other hand, for $T = 2\pi n$ we
have
$T$-periodic classical solutions for any initial value of $\theta$
and
$\varphi$ and the critical point set of the classical action
coincides   with the
classical phase space $S^2$.

{}From this example we can abstract the following generic properties:
The critical points of the vector field (\ref{lsvf}) {\it i.e.}
classical solutions
\be
\X^\mu_S ~=~ {\dot x}^\mu - \omega^{\mu\nu} \partial_\nu H ~=~ 0
\label{lscs}
\ee
with the boundary conditions $x^\mu (T) = x^\mu (0) = x^\mu_0$ form a
submanifold $\M_0$ of $\M$, the {\it moduli space} of classical
solutions. For
the class of Hamiltonians we consider {\it i.e.} Hamiltonians that
determine
the action of a circle on the phase space this moduli space can be
characterized as follows:

\vskip 0.2cm
- There are in general {\it discrete} values of $T$ for which the
classical
solutions (\ref{lscs}) admit nontrivial periodic solutions $x^\mu (0)
=   x^\mu(T) $ for
any initial condition $x^\mu(0) = x^\mu_0$ in a compact submanifold
$\M_0$ of
$\M$. In the example above with $T=2\pi n$, this submanifold
coincides with the
original phase space $\M$.

\vskip 0.2cm

- For {\it generic} values of $T$ the periodic solutions with
$x^\mu(0) =
x^\mu(T)$ can only exist if $x^\mu (0) = x^\mu_0$ is a point on the
critical
submanifold of $H$. In particular, the classical equations of motion
reduce to
\[
{\dot x}^\mu ~=~ \omega^{\mu\nu}\partial_\nu H ~=~ 0
\]
and consequently the moduli space $\M_0$ of $T$-periodic solutions in
this case
coincides with the critical point set of $H$, which is a submanifold
of $\M$.

\vskip 0.2cm
In order to localize (\ref{dhls2}), we need to construct an
appropriate   $\psi$.
For this we lift the $H$-invariant metric $g_{\mu\nu}$ on $\M$ to the
loop
space and define
\[
\psi ~=~ \frac{\lambda}{2} g_{\mu\nu} \X^\mu_S \eta^\nu
\]
As a consequence of (\ref{ldm}) the Lie-derivative condition
(\ref{ldls}) is   satisfied.
The evaluation of the $\lambda \to \infty$ limit follows
closely (that in section 3.: We write the loop space variable $x^\mu
(t)$ as
\[
x^\mu (t) ~=~ {\hat x}^\mu (t) + \delta x^\mu (t)
\]
where ${\hat x}^\mu(t)$ solves the classical equations of motion
\[
\X^\mu_S({\hat x}^\mu) ~=~ \partial_t \hat x^\mu -
\omega^{\mu\nu}(\hat x)
\partial_\nu H(\hat x) ~=~ 0
\]
and $\delta x^\mu (t)$ denotes the fluctuations. Similarly, we define
\[
\eta^\mu (t) ~=~ {\hat \eta}^\mu(t) + \delta \eta^\mu (t)
\]
where the ${\hat \eta}^\mu(t)$ are zeroes of the loop space
Riemannian
momentum map
\[
\Omega_{\mu\nu} ~=~ \frac{1}{2}\{ {\delta \over \delta x^\mu }
(g_{\nu\rho} \X^\rho_S) - {\delta \over \delta x^\nu }
(g_{\mu\rho} \X^\rho_S) \}
\]

\[
\Omega_{\mu\nu}(\hat z) {\hat \eta}^\nu ~=~ 0
\]
In particular this implies that the ${\hat \eta}^\mu$ are {\it Jacobi
fields},
{\it i.e.} satisfy the fluctuation equation
\[
[\delta_\nu^\mu \partial_t - \partial_\nu \X^\mu_S(\hat x) ] \ {\hat
\eta}^\nu
{}~=~ 0
\]

We define the loop space measure in (\ref{dhls2}) by
\[
[dx][d\eta] ~=~ d {\hat x}^\mu (t) d{\hat \eta }^\mu (t) \prod
d\delta x^\mu
(t) d\delta \eta^\mu (t)
\]
and introduce the path integral change of variables
\[
x^\mu (t) ~\to~ {\hat x}^\mu(t) + \frac{1}{\sqrt{\lambda}} \delta
x^\mu (t)
\]
\[
\eta^\mu (t) ~\to~ {\hat \eta}^\mu(t) + \frac{1}{\sqrt{\lambda}}
\delta
\eta^\mu (t)
\]
The corresponding Jacobian in the path integral measure is one.
We are interested in
the $\lambda \to \infty$ limit. By generalizing
the computation in section 3. to the loop space, we find that in this
limit the path integral (\ref{dhls2}) reduces to an integral over the
moduli   space $\M_0$
of classical solutions,
\be
Z ~=~ \int d{\hat x} (t) d{\hat \eta }(t) \ { \exp
\{ i \int\limits_0^T \vartheta_\mu {\dot {\hat x}}^\mu - H(\hat x)
+ \frac{1}{2}{\hat \eta}^\mu \omega_{\mu\nu} {\hat \eta}^\nu \}
\over {\rm Pf} || \delta_\mu^\nu \partial_t - \Omega_\mu^\nu (\hat x)
-
\frac{1}{2}
{R^\nu}_{\mu\rho\sigma}(\hat x) {\hat \eta}^\rho {\hat \eta}^\sigma
|| }
\label{lsdh2}
\ee
Here the Pfaffian is evaluated over the fluctuation modes $\delta
x^\mu$, and
${R^\mu}_\nu$ is the Riemannian curvature of the metric $g_{\mu\nu}$
evaluated
on the classical solution ${\hat x}^\mu(t)$. Notice in particular,
that the
measure in ( ) is an invariant measure over the moduli space of
classical
solutions which is itself a symplectic manifold.

In the limit, where we assume that the solutions to (\ref{lscs}) are
isolated and nondegenerate (for example if the critical points of $H$
itself
are isolated and nondegerate and if the period $T$ is such that the
boundary
condition $x^\mu(0) = x^\mu(T)$ only admits
constant loops as solutions to the classical equations of motion) we
can
further reduce (\ref{lsdh2}) to
\be
Z ~=~ \sum_{dS=0} { \exp \{ i S(\hat x) \} \over \sqrt{ \det ||
\delta_{\mu\nu} S || } }
\label{zds}
\ee

In conclusion, we have here reduced the path integral to an
integral
over the moduli space of classical solutions. In general this moduli
space has
a complicated, $T$ dependent structure. We shall now proceed to
derive an
alternative integration formula for (\ref{dhls}) which is applicable
independently of
the structure of the moduli space of classical solutions. It also has
the
advantage, that it can be directly generalized to loop space
equivariant Morse
theory, as we shall see in the following sections.

We select
\[
\psi ~=~ \frac{1}{2} g_{\mu\nu} {\dot x}^{\mu} \eta^\nu
\]
As a consequence of (\ref{ldm}) the condition (\ref{ldls}) is satisfied
and the corresponding action (\ref{dhls2}) is
\[
S ~=~ \int  \frac{1}{2} g_{\mu\nu} \dot x^\mu
\dot x^\nu + (\vartheta_{\mu} - \frac{1}{2} g_{\mu\nu}{\cal
X}_{H}^{\nu})
\dot x^\mu - H - \frac{1}{2} \eta^\mu ( g_{\mu\nu} \partial_{t} +
g_{\nu\sigma} \dot x^\rho \Gamma^{\sigma}_{\mu\rho} ) \eta^\nu +
\frac{1}{2}
\eta^\mu
\omega_{\mu\nu} \eta^\nu
\]
where $\Gamma^{\sigma}_{\mu\rho}$ is again the (metric) Levi-Civita
connection.
By the $\psi$ independence, the path integal ( \ref{dhls2}) remains
invariant under
the scaling $g_{\mu\nu} \to \lambda g_{\mu\nu}$ of the metric.  If
we evaluate it
in the $\lambda \to \infty$ limit we find \cite{oma6} that the result
is an   ordinary
integral over the classical phase space $\M$,
\[
Z  ~=~ \int dx d\eta \ \exp\{ -i T (H + \frac{1}{2} \eta^\mu
\omega_{\mu\nu} \eta^\nu) \} { 1 \over \sqrt{ det ||
{\delta^{\mu}}_{\nu}
\partial_{t} - \frac{1}{2} ({\Omega^{\mu}}_{\nu} +
{R^{\mu}}_{\nu\rho\sigma
}\eta^\rho\eta^\sigma) || } }
\]
We evaluate the determinant using {\it e.g.} $\zeta$-function
regularization.
The result is
\be
Z  ~=~ \int \! dx d\eta \ \exp\{ -i T (H +
\frac{1}{2} \eta^\mu  \omega_{\mu\nu} \eta^\nu ) \}
\cdot {\hat A} \left[ {T \over 2} ({\Omega^{\mu}}_{\nu} +
{R^{\mu}}_{\nu\rho\sigma}\eta^\rho\eta^\sigma ) \right]
\label{lsdh3}
\ee
where
\[
{\hat A} \left[ {T \over 2} ({\Omega^{\mu}}_{\nu} +
{R^{\mu}}_{\nu\rho\sigma}\eta^\rho\eta^\sigma ) \right] ~=~
\sqrt{ \det \left[ { \frac{T}{2} ({\Omega^{\mu}}_{\nu} +
{R^{\mu}}_{\nu\rho\sigma}\eta^\rho\eta^\sigma) \over
\sinh [ \frac{T}{2} ({\Omega^{\mu}}_{\nu} +
{R^{\mu}}_{\nu\rho\sigma}\eta^\rho\eta^\sigma) ] } \right] }
\]
is the equivariant $\hat A$-genus. This is our final integration
formula for
the path integral (\ref{dhls}), in terms of equivariant characteristic
classes   on the
classical phase space $\M$. Notice in particular, that here we
integrate over
the entire phase space $\M$  and not only over the moduli space
$\M_0$ of
classical solutions as in (\ref{lsdh2}), which is in general a
$T$-dependent submanifold of $\M$.

Since the equivariant $\hat A$-genus and the equivariant Chern class
are both
equivariantly closed with respect to $d_H$, we can further reduce
(\ref{lsdh3})   to the
critical point set $\M_0$ of $H$ by repeating the steps that led
to (\ref{dcdh}). For
this, instead of (\ref{lsdh3}) we consider the more general integral
\be
Z_\psi  ~=~ \int \! dx d\eta \ {\hat A} \left[ {T \over 2}
({\Omega^{\mu}}_{\nu} +
{R^{\mu}}_{\nu\rho\sigma}\eta^\rho\eta^\sigma ) \right]
\cdot \exp\{ -i T (H +
\frac{1}{2} \eta^\mu  \omega_{\mu\nu} \eta^\nu ) + d_H \psi \}
\label{lsdh4}
\ee
If $\psi$ is in the invariant subspace of $\cL_H$, according to our
standard
arguments (\ref{lsdh3}) and (\ref{lsdh4}) coincide. If we select
(\ref{psi}) and repeat the steps in section 3. we find that
(\ref{lsdh3}) localizes to the following integral over the critical
point set $\M_0$ \[
Z ~=~ \int\limits_{\M_0} dx d\eta \ {\hat A} \left[ {T \over 2}
({\Omega^{\mu}}_{\nu} +
{R^{\mu}}_{\nu\rho\sigma}\eta^\rho\eta^\sigma ) \right]
\cdot { \exp\{ -i T (H +
\frac{1}{2} \eta^\mu  \omega_{\mu\nu} \eta^\nu ) \} \over
{\rm Pf} ( \Omega_{\mu\nu} + R_{\mu\nu\rho\sigma} \eta^\rho
\eta^\sigma ) }
\]
of the classical Hamiltonian $H$, which generalizes (\ref{dcdh}) to
path integrals. In particular, if the critical   point set
$\M_0$ of $H$ is isolated and nondegenerate this reduces further to
the following generalization of (\ref{cdh}),
\[
Z ~=~ \sum\limits_{dH=0} {\hat A} \left[ {T \over 2}
{\Omega^{\mu}}_{\nu}
\right]
\cdot { \exp\{ -i T H \} \over
{\rm Pf} ( \Omega_{\mu\nu} ) }
\]


\vskip 1.0cm
\noindent
{\bf 7. Classical Morse theory and path integrals}
\vskip 0.7cm

We shall now proceed to generalize the loop space localization
methods of the
previous section to path integrals that are defined over the
supermanifold $S^*\M$. In the present section we shall consider the
infinite dimensional Matthai-Quillen formalism described in
\cite{mb} to derive some   standard
results in nondegenerate Morse theory. In the following section we
then
generalize these results to the equivariant and degenerate cases.

The various quantities constructed in section 4. can be directly
lifted to the
loop space $L(S^*\M)$ over $S^*\M$.  In particular, we define the
loop space
equivariant exterior derivative (in the following time integrals are
implicit)
\be
Q_t ~=~ \eta^\mu \frac{\partial}{\partial x^\mu} + p_\mu
\frac{\partial}{\partial \bar\eta_\mu} + \dot x^\mu i_\mu +
\dot{\bar\eta}_\mu \pi^\mu
\label{lsted1}
\ee
which is equivariant with respect to the natural action of the circle
$S^1$: $x^\mu(t) \to x^\mu (t + \tau)$ {\it etc.} on $L(S^*\M)$. The
corresponding Lie
derivative which acts on the exterior algebra over $L(S^* \M)$ is
\be
\L_t ~=~ Q_t^2 ~=~ \dot x^\mu \frac{\partial}{\partial x^\mu} + \dot
{\bar\eta}_\mu \frac{\partial}{\partial \bar\eta_\mu}  + \dot
\eta^\mu i_\mu +
\dot p_\mu \pi^\mu  ~\equiv~ {\partial \over \partial t}
\label{lsled}
\ee
We again introduce the conjugation (\ref{esed2}), (\ref{cfphi}) which
yields for  the equivariant exterior derivative (\ref{lsted1})
$$
Q_t ~\longrightarrow~ e^{\Phi} Q_t e^{-\Phi} ~=~ \eta^\mu
\frac{\partial}{\partial x^\mu} + (   p_\mu +
\Gamma^\rho_{\mu\nu} \eta^\nu \bar\eta_\rho )
\frac{\partial}{\partial
\bar\eta_\mu}
$$
$$
+ \dot x^\mu i_\mu +  ( \dot {\bar\eta}_\mu - \dot x^\nu
\Gamma^\rho_{\mu\nu}
\bar\eta_\rho +
\Gamma^\rho_{\mu\nu}p_\rho \eta^\nu - \frac{1}{2}
{R^\rho}_{\mu\nu\sigma}
\eta^\sigma \eta^\nu \bar\eta_\rho) \pi^\mu
$$
while the Lie derivative (\ref{lsled}) remains intact.

We are interested in deriving localization formulas for the following
path
integral over $L(S^*\M)$
\be
Z ~=~ \int [dx] [d\eta] [dp] [d\bar\eta] \exp\{ i \int Q_t \psi \}
\label{lstei}
\ee
which is of the standard form of  (cohomological) topological path
integral \cite{bbrt}. According to our general arguments, (\ref{lstei})
is invariant under local variations of $\psi$  provided these
variations   are in
the subspace
\be
\L_t \psi ~=~ 0
\label{lstldc}
\ee
But as a consequence of (\ref{lsled}) {\it any} functional $\psi$ which
is {\it single}-valued over the loops satisfies (\ref{lstldc}) since
\[
\L_t \psi ~=~ \int\limits_{0}^{T} dt \ \partial_t \psi ~=~
\psi(T) -
\psi(0)
\]

We select
\[
\psi ~=~ \frac{\lambda}{2} g_{\mu\nu} \dot x^\mu \eta^\nu  +
g^{\mu\nu} p_\mu
\bar\eta_\nu
\]
where $g_{\mu\nu}$ is a metric tensor on the original phase space
$\M$. This
gives for the action in (\ref{lstei})
\[
S ~=~ \int Q_t \psi ~=~ \int \frac{\lambda}{2} g_{\mu\nu} {\dot x}^\mu
{\dot   x}^\nu
+ \frac{\lambda}{2} \eta^\mu ( g_{\mu\nu} \partial_t + g_{\nu\sigma}
{\dot
x}^\rho
\Gamma^\sigma_{\rho\mu} ) \eta^\nu + \half g^{\mu\nu} p_\mu p_\nu
\]
$$
- \ \half {R^{\rho\mu}}_{\sigma\nu} \eta^\nu \eta^\sigma
\bar\eta_\rho
\bar\eta_\mu + \half (\bar\eta_\kappa g^{\kappa\mu}) ( g_{\mu\nu}
\partial_t -
g_{\nu\sigma} {\dot x}^\rho
\Gamma^\sigma_{\rho\mu} ) (g^{\nu\xi} \bar\eta_\xi )
$$
According to our standard arguments the corresponding path integral
is independent of $\lambda$, and we evaluate it in the $\lambda
\to\infty$ limit. For this we introduce the following local
coordinates on the loop space,
\ba
x^\mu(t) & = & x^\mu_{0} + x^\mu_{t}
\nonumber \\
\eta^\mu (t) & = & \eta^\mu_{0} + \eta^\mu_{t}
\nonumber \\
p_\mu (t) & = & p_{\mu 0} + p_{\mu t}
\nonumber \\
\bar\eta_{\mu}(t) & = & \bar\eta_{\mu 0} +  \bar\eta_{\mu t}
\label{lssov2}
\ea
with $x^\mu_{0} ,\ \dots \ \bar\eta_{\mu 0}$ the constant modes of
$x^\mu(t),
\ \dots \ \bar\eta^{\mu}(t)$ and $x^\mu_{t}, \  \dots \
\bar\eta^{\mu}_{t}$ the
$t$-dependent fluctuation modes. We define the path integral measure
by
\[
[dx][d\eta] [dp][d\bar\eta] ~=~ dx^\mu_{0} d\eta^{\mu}_{0} dp_{\mu 0}
d\bar\eta_{\mu 0}
\prod\limits_{t} dx^\mu_{t} d\eta^{\mu}_{t} dp_{\mu t} d\bar\eta_{\mu
t}
\]
and introduce the change of variables
\ba
x^\mu_{t} & \rightarrow & \frac{1}{\sqrt{\lambda}}x^\mu_{t}
\nonumber \\
\eta^{\mu}_{t} & \rightarrow & \frac{1}{\sqrt{\lambda}} \eta^{\mu}_{t}
\label{lscov2}
\ea
At least formally, the corresponding Jacobian is trivial. In
the $\lambda \to \infty$ limit we can evaluate the path integrals
over all fluctuation modes and the ordinary integrals over the
constant modes $p_{\mu 0}$ and $\bar\eta_{\mu 0}$. In this way we
find that
(\ref{lstei}) reduces to the integral of the Pfaffian of the curvature
two-form over the constant modes $x^\mu_0$ and $\eta^\mu_0$,
\be
Z ~=~ \int dx d\eta \ {\rm Pf}( R_{\mu\nu\rho\sigma \eta^\rho
\eta^\sigma} ) ~=~ \chi (M)
\label{tiph}
\ee
{\it i.e.} the path integral (\ref{lstei}) yields the Euler number of
the phase space $\M$.

In order to relate (\ref{lstei})  to the non-degenerate version of
the   Poincar\'e-Hopf
theorem, we introduce an arbitrary smooth vector field on $\M$ with
components
$V^\mu$ such that its zeroes are isolated and nondegenerate. (Notice
that {\it
e.g.} in \cite{mb} only gradient vector fields were considered.) We
then   select
\[
\psi ~=~ \frac{\lambda}{2} g_{\mu\nu} \dot x^\mu \eta^\nu \ + \ (\dot
x^\mu +
V^\mu)
\bar\eta_\mu
\]
This yields for the action in (\ref{lstei})
\[
S ~=~ \int Q_t \psi
\]
\[
=~ \int \frac{\lambda}{2} g_{\mu\nu} {\dot x}^\mu {\dot x}^\nu +
\frac{\lambda}{2} \eta^\mu ( g_{\mu\nu} \partial_t + g_{\nu\sigma}
{\dot
x}^\rho
\Gamma^\sigma_{\rho\mu}) \eta^\nu + p_\mu( {\dot x}^\mu + V^\mu) +
\bar\eta_\mu
( \delta_\nu^\mu
\nabla_t + \nabla_\nu V^\mu )\eta^\nu
\]
We again introduce the change of variables (\ref{lscov2}) for the
fluctuation modes. In the $\lambda \to \infty$ limit the integral
over $p_\mu(t)$ yields a $\delta$-function that localizes the
path integral over $x^\mu(t)$ to the zeroes of $V^\mu$. The remaining
path integrals are Gaussians, and evaluating these we
obtain
\be
Z ~=~ \sum\limits_{dV = 0} {\rm sign}( \det || \nabla_\mu V^\nu || )
\label{tigb}
\ee
Combining (\ref{tiph}) and (\ref{tigb}) we then have the familiar Morse
theory relation   between
the Poincar\'e-Hopf and Gauss-Bonnet-Chern theorems,
\be
\sum\limits_{dV = 0} {\rm sign}( \det || \nabla_\mu V^\nu || ) ~=~
\int dx d\eta \ {\rm Pf} ( R_{\mu\nu\rho\sigma} \eta^\rho\eta^\sigma )
\label{phgbc2}
\ee

We note that a generalization of (\ref{phgbc2}) to the degenerate case
can be derived by directly generalizing the computations in the
previous sections.


\vskip 1.0cm
\noindent
{\bf 8. Equivariant Morse theory in loop space}
\vskip 0.7cm

In order to derive path integral versions of the equivariant
Poincar\'e-Hopf
and Gauss-Bonnet-Chern theorems we need a loop space version of the
equivariant
exterior derivative $Q_\X$ in (\ref{qchi}). For this, we combine
(\ref{qchi}) and (\ref{lsted1}) to the following equivariant exterior
derivative on the loop space  $L(S^*\M)$,
\be
Q_S ~=~ d + i_S ~=~ \eta^\mu \frac{\partial}{\partial x^\mu} + p_\mu
\frac{\partial}
{\partial \bar\eta_\mu} + (\dot x^\mu - \X^\mu_H ) i_\mu + (\dot
{\bar\eta}_\mu
-
\partial_\mu \X^\nu_H \bar\eta_\nu) \pi^\mu
\label{qsils}
\ee
Notice that the zeroes of the $i_\mu$-components of the loop space
vector field in (\ref{qsils}) yield the equations of motion
$$
\dot x^\mu - \X^\mu_H ~=~ 0
$$
for the classical action
$$
S ~=~ \int \vartheta_\mu {\dot x}^\mu - H
$$
while the zeroes of the $\pi^\mu$-components of the vector field in
(\ref{qsils})
determines the Jacobi equation,
$$
(\delta^\mu_\nu \partial_t - \partial_\nu \X^\mu_H ) \bar\eta_\mu ~=~
0
$$
We again assume that
$$
\cL_H \Gamma_\mu ~=~ 0
$$
and introduce the conjugation (\ref{esed2}) which yields for
(\ref{qsils})
\[
Q_S ~\longrightarrow~
e^{\Phi} Q_S e^{-\Phi}
{}~=~ \eta^\mu \frac{\partial}{\partial x^\mu} + (p_\mu +
\Gamma^\rho_{\mu\nu}\eta^\nu \bar\eta_\rho ) \frac{\partial}{\partial
\bar\eta_\mu} + (\dot x^\mu - \X^\mu_H) i_\mu
\]
\[
+ \ \{ \Gamma^\rho_{\mu\nu} p_\rho \eta^\nu - \frac{1}{2}
{R^\rho}_{\mu\sigma\nu} \eta^\nu \eta^\sigma \bar\eta_\rho
- (\dot x^\nu - \X^\nu_H) \Gamma^\rho_{\mu\nu}\bar\eta_\rho + (
\delta^\rho_\mu
\partial_t
- \partial_\mu \X^\rho_H )\bar\eta_\rho \} \ \pi^\mu
\]
The pertinent conjugated Lie derivative is a linear combination
of (\ref{esld1}) and (\ref{lsled}),
\be
\L_S ~=~ \partial_t + \L_\X ~=~  \partial_t + \X^\mu_H
\frac{\partial}{\partial
x^\mu}
+ \eta^\mu \partial_\mu \X^\nu_H i_\nu - \partial_\mu\X^\nu_H
\bar\eta_\nu \frac{\partial}{\partial \bar\eta_\mu} - p_\nu
\partial_\mu  \X^\nu_H   \pi^\mu
\label{lsils}
\ee
and it determines the action of the vector field in
{\ref{qsils}) on the exterior algebra over the loop space $L(S^* \M)$.

We are interested in deriving localization formulas for the path
integral
\be
Z ~=~ \int [dx][dp][d\eta][d\bar\eta] \exp\{ i \int \vartheta_\mu
\dot
x^\mu - H + \half \eta^\mu \omega_{\mu\nu} \eta^\nu + Q_S\psi \}
\label{lsfpi}
\ee
Since (\ref{lsils}) is a linear combination of time
translation and (\ref{esld3}), we conclude that
{\it any generally covariant functional $\psi$ which is single valued
in the
loop space satisfies the $\psi$-independence condition}
\be
\L_S \psi ~=~ 0
\label{lsfldc}
\ee

We shall first derive an interpretation of (\ref{lsfpi}) corresponding
to the Poincar\'e-Hopf theorem in the case where the critical points of
the   action $S$
are isolated and nondegenerate. As we have explained in section 6.
this can be
the case for example if the period $T$ is properly selected and the
critical
point set of the Hamiltonian $H$ is isolated and nondegenerate.

We first introduce the functional
\be
\psi_1 ~=~ g^{\mu\nu} p_\mu \bar\eta_\nu
\label{psi1}
\ee
where $g_{\mu\nu}$ is again a metric tensor on the original phase
space $\M$
which is Lie-derived by the Hamiltonian vector field of $H$,
\[
{\cal L}_H g ~=~ 0
\]
As a
local and
generally covariant quantity (\ref{psi1}) then satisfies the
Lie-derivative   condition (\ref{lsfldc}) . A
direct computation yields for the last term in the action in
(\ref{lsfpi}) \[
Q_S \psi_1 ~=~ g^{\mu\nu}p_\mu p_\nu - \half {R^\rho}_{\mu\sigma\nu}
\eta^\nu
\eta^\sigma \bar\eta_\rho
g^{\mu\kappa}\bar\eta_{\kappa}
\]
\[
- \bar\eta_\mu(g^{\mu\nu} \partial_t +
\partial_\rho \X^\mu_H \
g^{\rho\nu})\bar\eta_\nu
+ \bar\eta_\mu g^{\mu\sigma}({\dot x}^\rho  -
\X^\rho_H
\Gamma^\nu_{\rho\sigma}) \Gamma^\nu_{\rho\sigma} \bar\eta_\nu
\]

Next, we introduce
\be
\psi_2 ~=~ \frac{\lambda}{2} g_{\mu\nu} (\dot x^\mu - \X^\mu_H)
\eta^\nu
\label{psi2}
\ee
where $\lambda$ is a parameter. This functional is also generally
covariant and
single valued on the loop space, and consequently it satisfies the
condition (\ref{lsfldc}). Explicitly, we find for the last term
in (\ref{lsfpi})
\[
Q_S\psi_2 ~=~ \frac{\lambda}{2} g_{\mu\nu} ({\dot x}^\mu - \X^\mu_H)
({\dot x}^\nu -  \X^\nu_H) +  \frac{\lambda}{2}
\eta^\mu \partial_\mu ( g_{\nu\rho}{\dot x}^\nu -  g_{\nu\rho}
\X^\nu_H) \eta^\rho
\]
We substitute both (\ref{psi1}) and (\ref{psi2}) in (\ref{lsfpi}) and
take the $\lambda\to \infty$ limit. By assuming that the solutions to
(\ref{lscs}) are   isolated
and nondegenerate and repeating the localization procedure
that led to (\ref{zds}), we find that in this limit (\ref{lsfpi})
reduces to
\[
Z ~=~ \sum\limits_{\delta S = 0} {\rm sign}( \det ||
\delta_{\mu\nu}S||) \exp
\{ i S \}
\]
where the sum is over all classical solutions (\ref{zds}) {\it i.e.}
critical points of the classical action $S$. This result shows, that
the path integral (\ref{lsfpi}) indeed can be related to an equivariant
version of the Poincar\'e-Hopf theorem in the loop space.

We shall now consider the following functional
\[
\psi_3 ~=~  \frac{\lambda}{2} g_{\mu\nu} {\dot x}^\mu \eta^\nu
\]
As a generally covariant and single valued loop space functional,
this
satisfies the Lie-derivative condition (\ref{lsfldc}). Explicitly, we
get for   the last
term in (\ref{lsfpi})
\[
Q_S \psi_3 ~=~  \frac{\lambda}{2} ({\dot x}^\mu - \X^\mu_H)
g_{\mu\nu} {\dot x}^\nu +  \frac{\lambda}{2} \eta^\mu ( g_{\mu\nu}
\partial_t + \dot   x^\rho
g_{\mu \sigma} \Gamma^\sigma_{\rho\nu}) \eta^\nu
\]
We then consider (\ref{lsfpi}) with
\[
\psi = \psi_1 + \psi_3
\]
We  introduce the change of variables (\ref{lscov2}) and  repeat the
steps that   led to (\ref{tiph}). In this way we find that as $\lambda
\to \infty$ (\ref{lsfpi}) localizes to   the
following integral over the original symplectic manifold $\M$,
\be
Z ~=~ \int\limits_\M dx d\eta \exp \{ -i T (H + \half \eta^\mu
\omega_{\mu\nu}
\eta^\nu )\} \ {\rm Pf} [ \half ({\Omega^\mu}_\nu  +
{R^\mu}_{\nu\rho\sigma}   \eta^\rho
\eta^\sigma ) ]
\label{iifp}
\ee
where ${\Omega^\mu}_\nu$ is again the Riemannian momentum map
(\ref{rmm}) and   ${R^\mu}_{\nu}$
is the curvature two-form on $\M$. In particular, we find that the
path
integral (\ref{lsfpi}) coincides with the finite
dimensional integral (\ref{egb2}).

Since the equivariant Chern class and the equivariant Pfaffian
are both equivariantly closed with respect to $d+i_H$ on the manifold
$\M$, we
can apply further localization to the integral (\ref{iifp}). If $\M_0$
again denotes   the
critical submanifold of $H$ in $\M$, following section 3. we then find
that (\ref{iifp})   reduces
further to
\[
Z ~=~ \int\limits_{\M_0} dxd\eta \ \exp\{ -iT(H + \omega) \}
{\rm Pf}[\half( \Omega^\mu_\nu + {R^\mu}_{\nu\rho\sigma}\eta^\rho
\eta^\sigma)]
\]
which can be further reduced to a sum over the critical points
of $H$
\[
Z ~=~ \sum\limits_{dH=0} e^{-iTH} {\rm sign} (\det ||
\partial_{\mu\nu} H || )
\]
provided these critical points are isolated and nondegenerate.


\vskip 1.0cm
\noindent
{\bf 9. Conclusions}
\vskip 0.7cm

In conclusion, we have shown how the Matthai-Quillen formalism can be
generalized by "equivariantizing" it with respect to a vector field
on a supersymmetric complex $S^*\M$. We have applied this
generalization to construct novel (path)integrals that
yield equivariant versions of the Poincar\'e-Hopf and Gauss-Bonnet-Chern
theorems in classical Morse theory. These (path)integrals are
naturally associated with integrable dynamical systems, suggesting that
for a large class of integrable models the quantum theory
could be formulated geometrically in terms of equivariant cohomology
on the classical moduli space of the theory.
In particular, our work indicates that there should be an intimate
relationship between cohomological topological field theories and
quantum integrable models.

\vskip 3.0cm

\noindent
{\bf Acknowledgements:} We thank A. Alekseev, V. Fock and A. Rosly for
discussions.

\pagebreak

\end{document}